\documentclass[conference]{IEEEtran}
\IEEEoverridecommandlockouts
\usepackage{cite}
\usepackage{amsmath,amssymb,amsfonts}
\usepackage{algorithmic}
\usepackage{graphicx}
\usepackage{textcomp}
\usepackage{xcolor}
\def\BibTeX{{\rm B\kern-.05em{\sc i\kern-.025em b}\kern-.08em
    T\kern-.1667em\lower.7ex\hbox{E}\kern-.125emX}}
\usepackage{tikz}
\usepackage{amsmath}
\usepackage{tikz}
\usepackage{amsmath}
\usepackage{url}
\usepackage{verbatim}
\usepackage{amsfonts}
\usepackage{graphicx}
\usepackage{textcomp}
\usepackage{xcolor}
\usepackage{breakurl}
\usepackage{multirow}
\usepackage{diagbox}
\usepackage{makecell}
\usepackage{booktabs}
\usepackage[normalem]{ulem}
\usepackage{enumitem}
\usepackage{fancyhdr} 
\usepackage{booktabs}
\usepackage{tikz}
\usepackage{colortbl}
\usepackage{balance}
\usepackage{CJKutf8}
\usepackage{threeparttable}
\usepackage{bbding}
\usepackage{hyperref} 

\usepackage{longtable}
\usepackage{supertabular,booktabs}
\usepackage{enumerate}

\newcommand{\FinalRevise}[1]{\textcolor{black}{#1}}

\usepackage{xcolor,pifont}
\newcommand*\colourcheck[1]{%
  \expandafter\newcommand\csname #1check\endcsname{\textcolor{#1}{\ding{52}}}%
}
\newcommand*\colourcross[1]{%
  \expandafter\newcommand\csname #1cross\endcsname{\textcolor{#1}{\ding{56}}}%
}
\colourcheck{black}
\colourcheck{red}

\colourcross{black}
\colourcross{green}

\newcommand{\tabincell}[2]{\begin{tabular}{@{}#1@{}}#2\end{tabular}}

\begin{document}

\title{Invisible Adversaries: A Systematic Study of Session Manipulation Attacks on VPNs}

    

\author{
    \IEEEauthorblockN{Yuxiang Yang\IEEEauthorrefmark{1}, Ao Wang\IEEEauthorrefmark{2}, Xuewei Feng\IEEEauthorrefmark{1}, Qi Li\IEEEauthorrefmark{3}\IEEEauthorrefmark{4},  and Ke Xu\IEEEauthorrefmark{1}\IEEEauthorrefmark{4}\Envelope}
    
    \IEEEauthorblockA{\IEEEauthorrefmark{1}Department of Computer Science and Technology \& BNRist, Tsinghua University}
    \IEEEauthorblockA{\IEEEauthorrefmark{2}School of Cyber Science and Engineering, Southeast University}
    \IEEEauthorblockA{\IEEEauthorrefmark{3}Institute for Network Sciences and Cyberspace \& BNRist, Tsinghua University, \IEEEauthorrefmark{4}Zhongguancun Lab}

   
   yangyx22@mails.tsinghua.edu.cn, wangao@seu.edu.cn, fengxw06@126.com, \{qli01, xuke\}@tsinghua.edu.cn \\
}

\maketitle

\begin{abstract}
Virtual Private Networks (VPNs) are widely used for censorship evasion and traffic protection. VPN users expect to be provided with adequate security protection, and at the same time not be affected by other users connected to the same VPN server, which can be illustrated as the non-interference property. 
However, in this paper, we have identified several vulnerabilities that violate this property, specifically within the connection tracking frameworks of VPN servers, stemming from shared resource misuse and insufficient validation of session state transitions. 
We present three session manipulation attacks targeting TCP and UDP traffic tunneled through VPNs. The attacker who only connects to the same VPN server can launch denial-of-service attacks, hijack TCP connections of other clients, or inject forged DNS responses into their queries. We evaluate these attacks against five popular connection tracking frameworks across different OSes and nine major commercial VPN providers. Experimental results reveal that all frameworks and eight providers are vulnerable to at least one of the attacks. We have responsibly disclosed our findings with countermeasures, resulting in 19 assigned CVEs/CNVDs and acknowledgments from the communities and providers.

\end{abstract}


\section{Introduction}

VPNs have emerged as a cornerstone in securing digital communication across the Internet~\cite{VPN-growth,Ramesh2022VPNalyzer}. By creating encrypted tunnels between devices and remote servers, VPNs shield information from prying eyes, ensuring data confidentiality, integrity, and authenticity. 
Nevertheless, according to the non-interference property~\cite{non-interference-property}, VPN servers should be properly configured to ensure that processes from different users remain isolated and do not interfere with one another.

Previous studies have highlighted critical VPN vulnerabilities, including cryptographic weaknesses~\cite{pptp-ccs98,ipsec-ccs15,ipsec-sec18}, endpoint misconfigurations~\cite{vpn-dns-hijack-pets15,vpn-configuration-imc16,vpn-ecosystem-imc18,VPN-alyzer-ndss22,vpn-routetable-sec23} and in/on-path adversaries exploiting features such as packet size patterns~\cite{sec21-Tolley}, DHCP behaviors~\cite{tunnel-vision}, or local IP manipulation~\cite{vpn-routetable-sec23}. 
Recently, Mixon-Baca et al.~\cite{MixonBaca-pets24} proposed attacks breaching non-interference to redirect traffic or de-anonymize VPN peers. Their work primarily focuses on port conflicts between VPN clients and the VPN server itself, and the feasibility of the attack on real-world VPNs remains unverified. 
Compared with previous works to hijack TCP~\cite{qian2012,gilad-fragmentation-tcp-2013,cao2016off,chen2018off,ccsfeng} and UDP~\cite{dns-Herzberg-tpds12,herzberg-fragmentation-dns-2013-cns,gilad-tcp/dns-hacking-2014,DNS-Man-2020,DNS-Man-2021,yangexploiting} sessions in plaintext environments, session-level vulnerabilities in VPN ecosystems remain underexplored. 
This leaves open questions about VPN session isolation guarantees in real-world deployments.

In this paper, we conduct an in-depth exploration of session-level vulnerabilities in VPN ecosystems. We identify critical flaws in the shared connection tracking table and the insufficient validation of session state transitions at VPN servers. Then we propose three session manipulation attacks targeting TCP and UDP traffic tunneled through VPNs, which can be launched by an off-path adversary who shares the same VPN server as the victim.
First, as the VPN server's public-facing source ports to a target server are limited, the attacker can exhaust all of them to launch a port exhausting DoS attack to the victim. 
Second, when the connection tracking framework of the VPN server adopts \textit{Port Preservation} when managing session entries, the attacker can deduce randomized source ports of the victim sessions by sending probe and verification packets with guessed ports and observing the destination of the verification packets. Furthermore, it can abuse the inferred source ports to launch a TCP hijacking attack or spoof responses to UDP-based DNS queries from other VPN users.
Third, the connection tracking frameworks may change session states when receiving crafted TCP \texttt{RST} packets without sufficiently verifying the sequence numbers. The attacker can remove others' session entries, and replace itself as the clients, thus totally hijacking the TCP sessions.

We verify these vulnerabilities and attacks on 5 popular connection tracking frameworks (i.e., Linux \texttt{Netfilter}, FreeBSD \texttt{PF}, \texttt{IPFW}, \texttt{IPFilter}, and \texttt{natd}) and 9 popular commercial VPN providers~\cite{top10vpn} with ethical guidelines. 
Our case studies on session DoS, HTTP injection, and DNS hijacking show that all frameworks and 8 providers are vulnerable to at least one of the attacks.
On average, an attacker can launch a DoS attack in 4 seconds with over 90\% success, and manipulate HTTP traffic within 64.11 seconds at a 66.7\% success rate. DNS hijacking is also feasible, with success rates ranging from 20\% to 70\%. These results highlight that breaching non-interference undermines VPN security guarantees.

Finally, we propose countermeasures that aim to disrupt the attack prerequisites and uphold non-interference at the core of VPN security.
We disclosed the vulnerabilities and attacks to Linux, FreeBSD, and affected VPN providers. Till now, we have received acknowledgments from Linux, FreeBSD, and six VPN vendors rewarded us under bug bounty programs. The other two vendors are still investigating them. 
In total, our findings have led to the assignment of 19 CVEs/CNVDs.

\noindent \textbf{Contributions}. Our main contributions are as follows:
\begin{itemize}[leftmargin=*,itemsep=2pt,topsep=0pt,parsep=0pt]
    \item We uncover new vulnerabilities at the VPN server due to the violation of the non-interference property, which can be exploited by malicious users to manipulate others' sessions.
    \item We verify the existence of the vulnerabilities against popular connection tracking frameworks and evaluate the session manipulation attacks against well-known VPN vendors. 
    \item We propose practical countermeasures and have responsibly disclosed the issues to the relevant organizations. We make our code publicly available at \url{https://github.com/yyxRoy/vpn-attacks} for ease of replication.
    
\end{itemize}

\section{Background and Threat Model}\label{sec:background}
\subsection{VPN Architecture and Connection Tracking}\label{sec:background-vpn-conntrack}

VPNs enable secure communication over public networks by establishing encrypted tunnels between client devices and remote servers.
In practice, modern VPN implementations usually create a virtual network interface (e.g., \texttt{tun0}) on the client side, which redirects traffic through a secure tunnel to the VPN server. All outbound packets are routed via this interface, encrypted, and then forwarded to the VPN server, which decrypts and relays them to the destination. To support IP address translation and session management, most VPN servers rely on connection tracking frameworks (e.g., Linux \texttt{Netfilter} or FreeBSD \texttt{ipfw}) to maintain per-session state. These frameworks act as stateful NAT middleboxes, rewriting packet headers by replacing the client's source IP with the server's public IP, and tracking session metadata (e.g., port, protocol, state, timeout) in a shared connection tracking table.

When an internal client \texttt{C} initiates a connection to a remote server \texttt{S}, the VPN server \texttt{V} will create a session entry that maps \texttt{C}'s internal address and port to \texttt{V}'s external ones. Consider \texttt{C} sending a TCP \texttt{SYN} packet with source port \texttt{c} to \texttt{S} listening on port \texttt{s} via \texttt{V}. The entry is recorded as:
\vspace{-1mm}
\begin{equation}\label{eq:syn_entry}	
\resizebox{0.91\hsize}{!}{$\begin{aligned}
\{[C:c \leftrightarrow V:c] \leftrightarrow [S:s],\,tcp=SYN\_SENT,\,timeout=120 s\}		
\end{aligned}$}
\vspace{-1mm}
\end{equation}
Depending on the port allocation strategy, \texttt{V} may either attempt \textit{Port Preservation} (reusing the same source port \texttt{c}) or apply \textit{Random Selection} (assigning a new external port). Port conflicts are resolved by selecting unused ports dynamically. 
As the TCP session progresses, the entry state transitions accordingly (e.g., from \textit{SYN\_SENT} to \textit{ESTABLISHED}) based on observed packets such as \texttt{SYN/ACK} and \texttt{ACK}. These transitions enable the framework to manage flows and reject unsolicited packets~\cite{rfc5382}.
Moreover, to accommodate asymmetric routing or middlebox failures, most connection tracking frameworks enable a \textit{loose state instantiation} mechanism by default, which permits the creation of a new \textit{ESTABLISHED} entry directly from incoming non-SYN packets (e.g., \texttt{PUSH/ACK}) without a three-way handshake, effectively restoring the session mapping with the sequence numbers provided in the packet.

\subsection{Threat Model}\label{sec:threat-model}

\begin{figure}[ht]
	\begin{center}
		\includegraphics[width=1\linewidth]{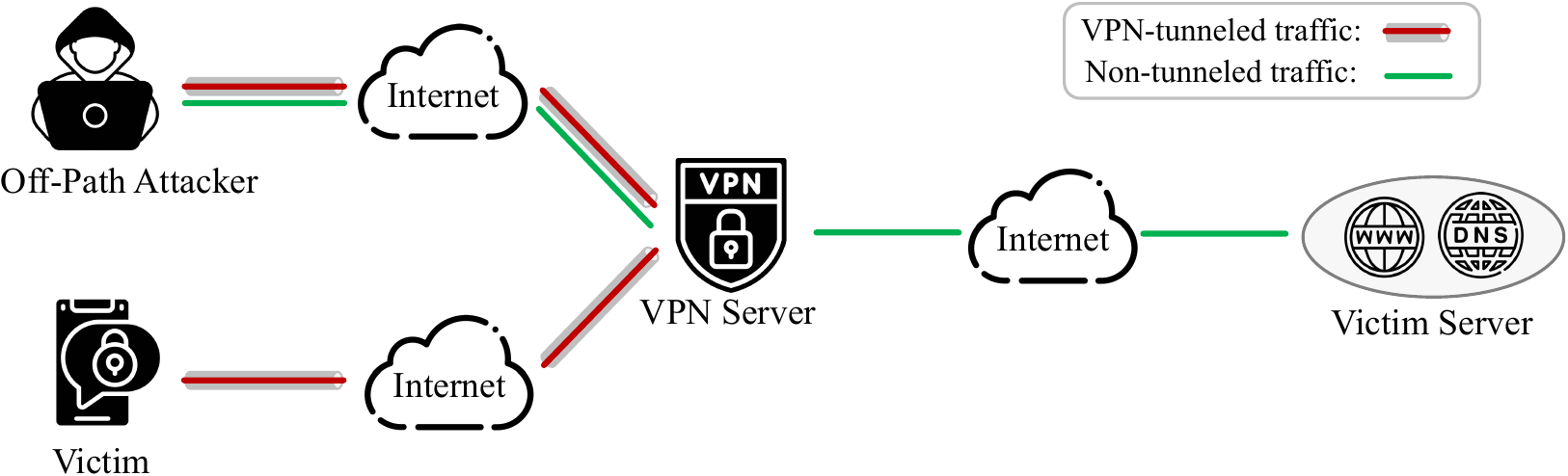}
		\caption{Threat model}
		\label{fig:threat-model}
	\end{center}
	\vspace{-2mm}
\end{figure}

Figure \ref{fig:threat-model} illustrates the threat model of our attacks.
The target server provides services by accepting requests to an open port (e.g., 53 for DNS, and 80 for HTTP) and returning the responses. Depending on different scenarios, the target server can be chosen as popular servers as previous works \cite{cao2016off,chen2018off,sec21-Tolley,ccsfeng,yangexploiting} (e.g., search engines, famous websites) or the default DNS resolvers shared by the VPN users~\cite{sec21-Tolley}.
The VPN server works as an intermediate to protect the users who connect to it, which will transfer packets traversing through it by maintaining a stateful connection tracking table.
The victim VPN user who has connected to the VPN server tries to communicate with the target server.
For instance, it may send a request periodically to the target server and await the responses from it (e.g., DNS lookup, HTTP requests).
%
The physically off-path attacker also connects to the VPN server and aims to manipulate the sessions of other users. It can send normal packets through the VPN tunnel (as the tunneled traffic in red color outlined in Figure \ref{fig:threat-model}) and is also capable of sending spoofed packets with the IP addresses of the target server outside of the tunnel (as those in green color). As reported in \cite{SAV-deployment,spoofer-caida}, lots of ASes in the world still do not validate packets with spoofed source addresses.
\FinalRevise{
The physically off-path attacker also connects to the VPN server and aims to manipulate the sessions of other users. Since it has full control over its devices, it can selectively route normal packets through the VPN tunnel via the virtual network interface (e.g., \texttt{tun0}) (as the tunneled traffic in red color outlined in Figure \ref{fig:threat-model}).
Simultaneously, we assume it is capable of sending spoofed packets directly to the VPN server's public IP address via the physical interface (e.g., \texttt{eth0}) (as the non-tunneled traffic in green color).
Recent studies~\cite{SAV-deployment,ccsfeng} demonstrate that IP spoofing remains widely feasible across the Internet.
}

As required by the non-interference property~\cite{non-interference-property}, the behavior of one user or process in a system should not interfere with that of other users or processes. 
VPN servers typically enforce client isolation policies to block peer-to-peer communication within the tunnel, preventing the attacker from scanning the virtual subnet to discover other active clients.
Consequently, the attacker initially has no knowledge of the victim's identity (e.g., the internal IP address assigned by the VPN or the client's real public IP) or the specific ephemeral source port allocated for its outgoing sessions.
However, we find that the shared resources of connection tracking frameworks of the VPN server violate this property, which allows a malicious VPN user to abuse the vulnerabilities to interfere with other users' sessions. Based on this threat model, we propose three practical session manipulation attacks, i.e., port exhausting DoS attack, TCP hijacking attack, and DNS hijacking attack.

\section{Session Manipulation Attacks }\label{sec:attack}

\subsection{Port Exhausting DoS Attack}\label{sec:4-sub-port-DoS-attack}

Once the attacker has connected to the VPN server, it shares the network resources of the VPN server with the other VPN users that connect to it. However, a notable drawback of the shared environment lies in the source port limitation when considering upper-level transport protocols.
Given that the source port field in TCP and UDP protocols is limited to 16 bits, a VPN server can use only 65,535 distinct source ports when initiating sessions to a target server. Consequently, once all these source ports are occupied by other sessions, users are rendered unable to initiate any outbound session to the target server, thus facilitating a port exhausting DoS attack.

\begin{figure}[h]
	\begin{center}
		\includegraphics[width=1\linewidth]{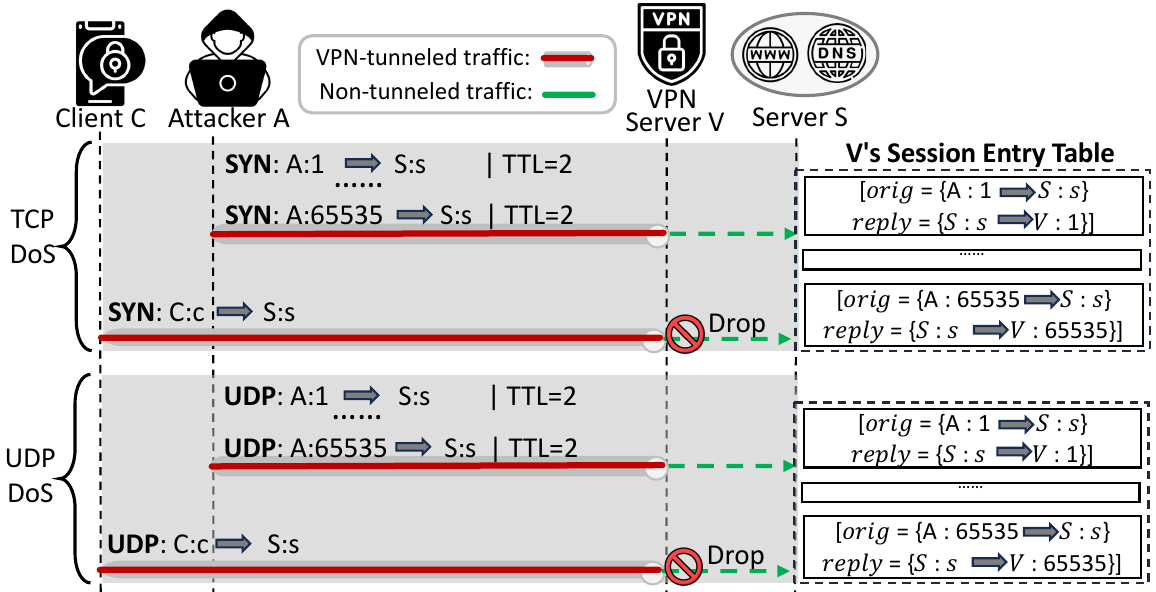}
		\caption{Denial of service by exhausting usable ports}
		\label{fig:port_dos}
	\end{center}
\end{figure}

Figure~\ref{fig:port_dos} depicts the steps of the attack. For a target server \texttt{S}, the attacker can launch the attack by sending outgoing packets to consume all of \texttt{V}'s available ephemeral ports for the connection 4-tuple. 
Take TCP protocol as an example, the attacker first sends 65,535 \texttt{SYN} packets with different source ports to \texttt{S} through the VPN tunnel. To prevent a large number of packets from arriving at \texttt{S}, the attacker can set the \texttt{TTL} of the \texttt{SYN} packet to a small number (e.g., 2), resulting in being dropped at the intermediate routers. \texttt{V} will create 65,535 session entries in the state of \textit{SYN\_SENT}, which will last for a certain period (e.g., 120 seconds by default in \texttt{Netfilter}).
Moreover, the attacker can continue flushing the entries' \textit{timeout} by sending outgoing packets again and again when they are closed to be cleaned. 

After that, other VPN users cannot visit the target server anymore, as there is no ephemeral port left for a new session. 
Furthermore, since all VPN clients share the same DNS resolvers in a well-configured VPN setting, the attacker can amplify the attack by occupying all source ports to the DNS resolvers in advance, making others unable to resolve domain names and resulting in a more serious DoS attack.
The attack clearly violates the non-interference property, where the attacker's actions disrupt other users' normal VPN usage. 

The attack is more stealthy than traditional NAT DoS techniques that exhaust the entire table or system resources~\cite{winemiller2012nat,slowdos-CANDAR-2018,slowdos-2022}, which often generate explicit logs (e.g., ``\textit{nf\_conntrack: table full, dropping packet}'' in Linux), making them easier to detect by network administrators. In contrast, our approach targets specific servers and selectively consumes ephemeral ports, producing no system logs and using fewer resources.
Setting low \texttt{TTL} values ensures packets are dropped before reaching the target, further reducing the detectability.

\noindent \textbf{Enhancing Attack Persistence.}
Instead of leaving entries in the \textit{SYN\_SENT} state, the attacker may spoof \texttt{SYN/ACK} and \texttt{ACK} packets to complete the TCP handshake, transitioning entries to the \textit{ESTABLISHED} state, which typically has a much longer timeout (e.g., 5 days in \texttt{Netfilter}). The same methods can be applied to the UDP protocol as well.

\subsection{TCP Hijacking Attack}\label{sec:attacks-TCP-Hijacking}

The procedures to perform this attack are as follows:


\subsubsection{Making Inferences about Active TCP Sessions}\label{4-subsec-inference-sessions}

\begin{figure}[ht]
	\begin{center}
		\includegraphics[width=0.95\linewidth]{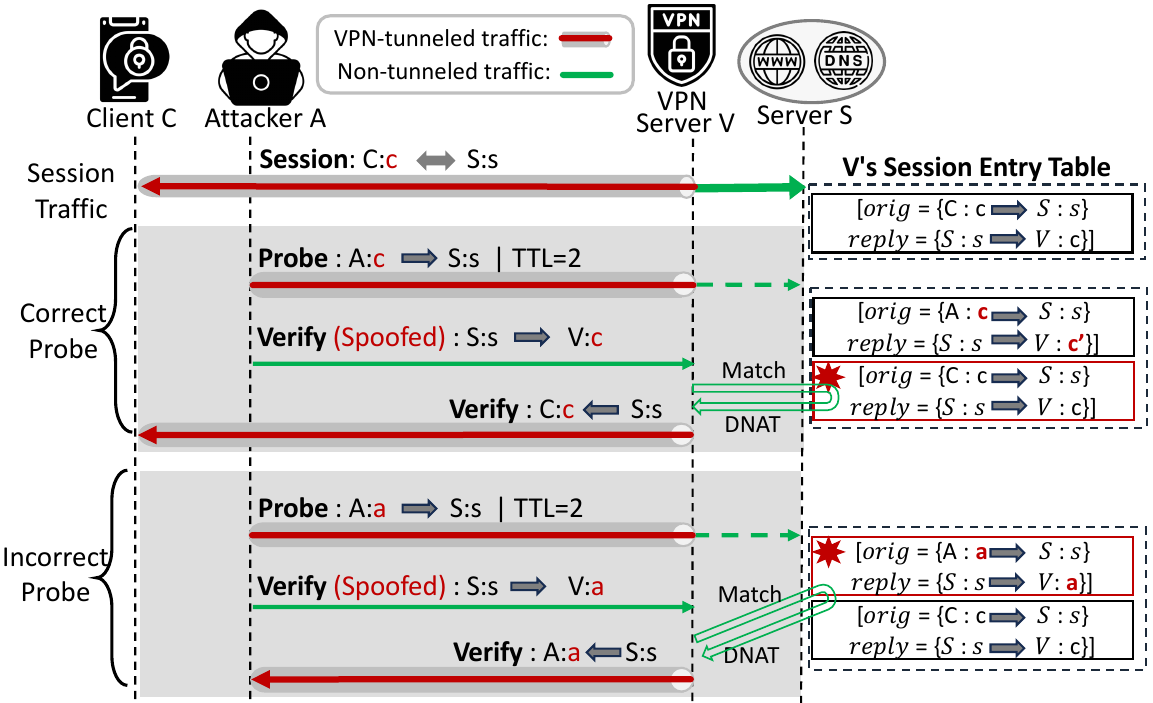}
		\caption{Inferring the source port of the victim session}
		\label{fig:connectioninference}
	\end{center}
\end{figure}

The attacker \texttt{A} aims to determine whether a client \texttt{C} is communicating with a server \texttt{S} on TCP port \texttt{s}, and if so, identify \texttt{C}'s source port \texttt{c}. If the VPN server adopts \textit{Port Preservation}, it maintains a session entry as:
\vspace{-1mm}
\begin{equation}\label{eq:client_entry}	
\{[C:c \leftrightarrow V:c] \leftrightarrow [S:s]\}	
\vspace{-1mm}
\end{equation}
The attacker \texttt{A} has to probe the ephemeral port space to determine the port used by \texttt{C} with the steps in Figure~\ref{fig:connectioninference}. There will be two cases:

\noindent a). Hitting the port \texttt{c}: First, \texttt{A} sends a \texttt{Probe} packet (i.e., \texttt{SYN}) packet from source port \texttt{c} to \texttt{S:s} via the VPN tunnel, i.e., \{\texttt{A:c$\rightarrow$S:s}, \texttt{SYN}\}. Due to port \texttt{c} already being used by \texttt{C}, the VPN server \texttt{V} maps the packet to a different external port \texttt{c'}, resulting in the entry:
\vspace{-1mm}
\begin{equation}\label{eq:attacker_right_entry}	
\{[A:c \leftrightarrow V:c'] \leftrightarrow [S:s]\}	
\vspace{-1mm}
\end{equation}
The attacker then send a \texttt{Verify} packet (i.e., spoofed \texttt{SYN/ACK}) impersonating \texttt{S} to \texttt{V} with port \texttt{c} outside the VPN tunnel, i.e., \{\texttt{S:s$\rightarrow$V:c}, \texttt{SYN/ACK}\}.
When the \texttt{SYN/ACK} arrives, it will match entry (\ref{eq:client_entry}) and will be forwarded to \texttt{C}. 

\noindent b). Hitting an unused port \texttt{a}: The attacker sends the \{\texttt{A:a$\rightarrow$S:s}, \texttt{SYN}\} packet through the VPN tunnel first. \texttt{V} will create an entry (\ref{eq:attacker_wrong_entry}) that keeps the source port, as there is no collision. Then, when the spoofed \{\texttt{S:s$\rightarrow$V:a}, \texttt{SYN/ACK}\} arrives at the VPN server, it will be forwarded back to the attacker according to entry (\ref{eq:attacker_wrong_entry}). 
\vspace{-1mm}
\begin{equation}\label{eq:attacker_wrong_entry}	
\{[A:a \leftrightarrow V:a] \leftrightarrow [S:s]\}
\vspace{-1mm}
\end{equation}
To reduce detectability, the attacker can set a low \texttt{TTL} for \texttt{Probe} packets, ensuring they do not reach the server. By scanning the full ephemeral port range, the attacker can infer all active sessions, violating the non-interference property.

\subsubsection{Obtaining \textit{SEQ} and \textit{ACK} Numbers}\label{4-subsec-obtain-seq}

After determining the source port \texttt{c}, the attacker attempts to obtain the sequence and acknowledgment numbers of the victim connection recorded in entry (\ref{eq:client_established_entry}). 
Earlier approaches primarily deduced these values by exhaustively exploring the entire 4G spectrum via side channels~\cite{ccsfeng,cao2016off,chen2018off,sec21-Tolley}, which is time-consuming and unreliable.
In this work, we find that vulnerabilities exist in connection tracking frameworks that change the state of the session entries when receiving TCP \texttt{RST} packets without sufficiently verifying the sequence numbers. 
Here we use \texttt{Netfilter} to illustrate our attack.
Other frameworks (e.g., \texttt{PF}, \texttt{IPFilter}) exhibit slight variations in state management and attack details (see Section~\ref{5-subsec:analysis-of-NAT-frameworks}). 
\vspace{-2mm}
\begin{equation}\label{eq:client_established_entry}	
\resizebox{0.91\hsize}{!}{$\begin{aligned}
\{[C:c \leftrightarrow V:c] \leftrightarrow [S:s],\,tcp=ESTABLISHED,\,timeout=432000 s\}
\end{aligned}$}
\vspace{-2mm}
\end{equation}

Before the patch~\cite{Netfilter-patch}, upon receiving an in-window (a value close to $2^{16}$ in our test) TCP \texttt{RST}, \texttt{Netfilter} will directly transfer the session state from \textit{ESTABLISHED} (432,000s timeout) to \textit{CLOSE} (10s timeout), which is very vulnerable to blind TCP reset attacks.
After the patch, \texttt{Netfilter} will keep the state of the session as \textit{ESTABLISHED}, decrease the timeout to 10 seconds, and wait for the challenge ACK from the endpoint (according to RFC 5961~\cite{rfc5961}) to restore the timeout to 300 seconds, which is much harder for attackers to abuse.


\begin{figure}[h]
	\begin{center}
		\includegraphics[width=0.95\linewidth]{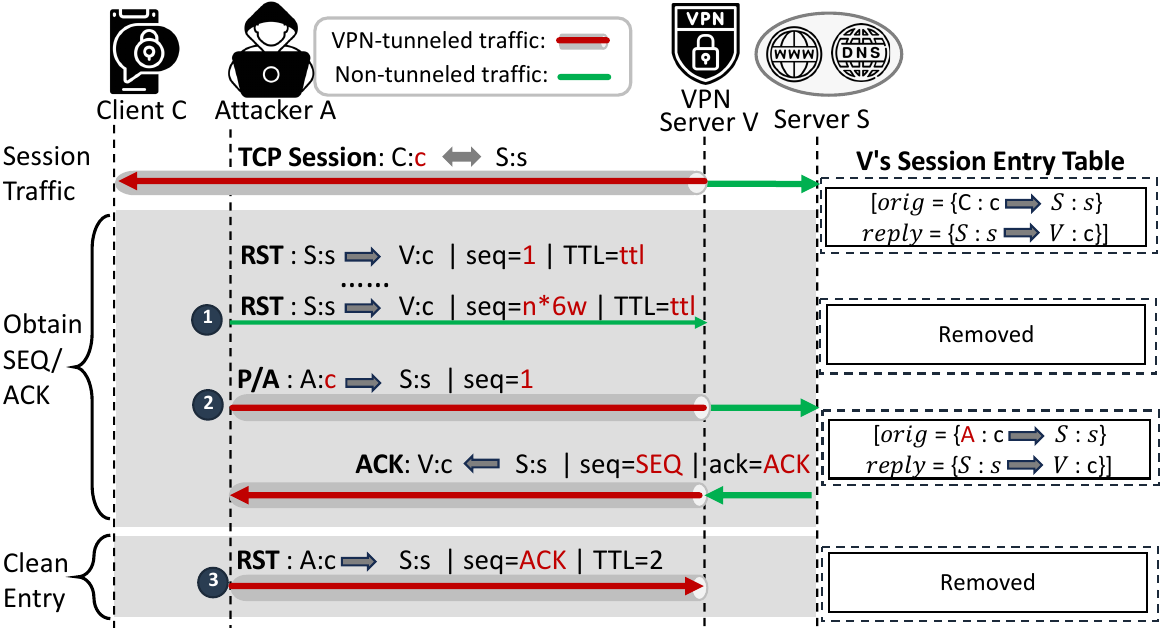}
		\caption{Obtaining SEQ and ACK numbers}
		\label{resetmapping}
	\end{center}
    	\vspace{-4mm}
\end{figure}

We propose a new method to bypass the verification as shown in Figure~\ref{resetmapping}.
\ding{182}As \texttt{Netfilter} will only check whether the \texttt{RST} sequence number is within the challenge ACK window, the attacker can remove the victim session entry by sending spoofed TCP \texttt{RST} packets with varying sequence numbers as the target server, i.e., \{\texttt{S:s$\rightarrow$V:a}, \texttt{RST}, \texttt{seq=i}, \texttt{TTL}= \texttt{ttl}\}, to the VPN server outside the VPN tunnel. We set the sequence number interval to 60000 to ensure that one of them will be located within the acceptable range, and the attacker only needs to send near 71k \texttt{RST} packets, which takes only seconds for modern machines.

To bypass the patch~\cite{Netfilter-patch}, the attacker can prevent triggering \texttt{challenge ACK} via controlling the \texttt{TTL} of the \texttt{RST} packets.
It can probe the hop distance to the VPN server with tools (e.g., Traceroute), and then specify it in the \texttt{RST} packets.
Upon arrival, the in-window \texttt{RST} packet will first incur the state transition of the victim session entry and then be dropped as the \texttt{TTL} value has decreased to 0, thus no \texttt{challenge ACK} from the victim client will be triggered anymore. 
The entry (\ref{eq:client_established_entry}) will remain in \textit{ESTABLISHED} while its expiration \textit{timeout} will decrease to 10 seconds, as shown in entry (\ref{eq:client_entry_after_RST}). After its timeout, the entry will be completely removed.
\vspace{-2mm}
\begin{equation}\label{eq:client_entry_after_RST}	
\resizebox{0.91\hsize}{!}{$\begin{aligned}
\{[C:c \leftrightarrow V:c] \leftrightarrow [S:s],\,tcp=ESTABLISHED,\,timeout=10 s\}		
\end{aligned}$}
\vspace{-2mm}
\end{equation}
\ding{183}Then the attacker replaces the victim client by constructing a new session entry (\ref{eq:attacker_entry_after_RST}) via sending a data packet with its own private IP address to the remote server with arbitrary sequence and acknowledgment numbers, e.g., \{\texttt{A:c$\rightarrow$S:s}, \texttt{PUSH/ACK}, \texttt{seq=1}, \texttt{ack=1}\}.
\vspace{-1.5mm}
\begin{equation}\label{eq:attacker_entry_after_RST}
\resizebox{0.91\hsize}{!}{$\begin{aligned}
\{[A:c \leftrightarrow V:c] \leftrightarrow [S:s],\,tcp=ESTABLISHED,\,timeout=300 s\}		
\end{aligned}$}
\vspace{-1.5mm}
\end{equation}
After translation by the VPN server, the packet is routed to the target server. From the perspective of the remote server, this packet matches the victim session and will incur an \texttt{ACK} packet, i.e., \{\texttt{S:s$\rightarrow$V:c}, \texttt{ACK}, \texttt{seq}=\textit{SEQ}, \texttt{ack}=\textit{ACK}\} back with the server's exact \textit{SEQ} and \textit{ACK}~\cite{rfc793}. 
When the \texttt{ACK} packet arrives at the VPN server, it will be translated and routed to the attacker (entry (\ref{eq:attacker_entry_after_RST})), allowing it to obtain the \textit{SEQ} and \textit{ACK} directly. 

\subsubsection{Hijacking Active TCP Sessions}\label{4-subsec-hijack-sessions}

With these values, the attacker can directly launch a TCP DoS attack by sending a TCP \texttt{RST} packet  to the remote server, or it can imitate the victim client to send requests to the remote server and hijack the responses.
\ding{184}In addition, to inject forged responses into the victim client, the attacker needs to first remove its session entry (\ref{eq:attacker_entry_after_RST}) by sending a TCP \texttt{RST} (Step 3 in Figure \ref{resetmapping}). It then continuously sends forged responses to the VPN server impersonating the target server outside the tunnel, i.e., \{\texttt{S:s$\rightarrow$V:c}, \texttt{PUSH/ACK}, \texttt{seq}=\textit{SEQ}, \texttt{ack}=\textit{ACK}\}. These packets are initially dropped silently due to the removed session entry. Once the victim client launches the next request and restores the entry, the forged responses will be forwarded to the victim client, thus achieving TCP injection.

\subsection{DNS Hijacking Attack}\label{sec:attacks-DNS-Hijacking}

According to \cite{sec21-Tolley} and our empirical studies, when a client is visiting a specific website on mainstream browsers, once the domain entry expires in the browser's DNS cache, ordinary user actions (e.g., loading in-page subresources or refreshing pages) can periodically trigger DNS resolution on the target domain name. For instance, after the DNS cache entry for a domain name (\texttt{a.com}) expires, accessing a resource within the web page (\texttt{a.com/resources}) dispatches a DNS request. We summarize the DNS cache behavior of four widely used browsers across different desktop OSes in Table~\ref{DNS_cache_timeout}. Overall, most browsers either follow the DNS response TTL or apply a short fixed caching window (e.g., 60\,s), which opens a recurring window for an attacker to race and inject spoofed DNS responses.
In contrast to previous DNS cache poisoning attacks targeting public DNS resolvers \cite{DNS-Man-2020,DNS-Man-2021}, as the VPN server does not provide a DNS cache, our DNS hijacking attack does not aim to contaminate caches, but to inject forged responses into victims' DNS queries as Tolley's work \cite{sec21-Tolley}. Although the impact may be relatively less severe compared to cache poisoning, considering the capabilities and threat model of an off-path attacker, this attack's effectiveness already demonstrates its harmful potential.

\begin{table}[h]
\centering
\caption{DNS Cache Behavior of Different Browsers.}
\label{DNS_cache_timeout}
\begin{threeparttable}
\scalebox{0.75}{
\begin{tabular}{cc|c|c} 
\bottomrule
\textbf{OS} & \textbf{Browser} & \textbf{Version} & \textbf{DNS Cache Timeout} \\ \hline
\multirow{3}{*}{Windows 11} & Chrome & 143.0.7499.170 & 60s \\
 & Edge & 143.0.3650.96 & 60s \\
 & Firefox & 146.0.1 & $\geq$ 660s \\ \hline
\multirow{4}{*}{macOS 26.1} & Chrome & 143.0.7499.170 & $\max(60s, \text{Response's TTL})$ \\
 & Edge & 143.0.3650.80 & Response's TTL \\
 & Safari & 21622.2.11.11.9 & $\approx \text{Response's TTL} \times 1.5$ \\
 & Firefox & 146.0.1 & 660s \\ \hline
\multirow{3}{*}{Ubuntu 22.04} & Chrome & 143.0.7499.192 &  Response's TTL \\
 & Edge & 143.0.3650.96 & Response's TTL \\
 & Firefox & 146.0.1 & 660s \\ 
\toprule
\end{tabular}
}
\end{threeparttable}
\end{table}

Besides identifying the correct TxID, the attacker must also determine whether the victim is visiting the target domain. We follow the similar steps as \cite{sec21-Tolley} that the attacker first performs TCP inference (Section~\ref{4-subsec-inference-sessions}) to detect long-lived TCP sessions to IPs associated with the target domain. If TCP sessions exist, the attacker can then ascertain that the victim client is likely accessing a website associated with the target domain. As it is quite possible for the browser to launch DNS requests of the domain periodically, the attacker can start to scan the source port range to launch a DNS hijacking attack on the target domain name by following the two steps:


\subsubsection{Making Inferences about Active DNS Sessions}\label{subsec:DNS_infer_port}

Similar to Section \ref{4-subsec-inference-sessions}, the attacker needs to determine the source port of the DNS request from the user to the target DNS resolver. The VPN server will maintain a session entry (e.g., entry (\ref{eq:client_entry})) for the user's DNS request. 
The attacker can scan the entire port number space following the same procedure in Figure \ref{fig:connectioninference}. The difference is that the \texttt{Probe} and \texttt{Verify} packets are UDP packets with the guessed source ports. After the probing and verifying process, the attacker can determine the source port \texttt{c} used by the victim's DNS request.

\subsubsection{Injecting DNS Responses via Brute Forcing TxIDs}\label{subsec:inject_dns_response}

DNS packets contain a 16-bit transaction ID (\texttt{TxID}), requiring the attacker to brute-force up to 65k possible values by sending spoofed responses to the inferred UDP session. This is feasible with modern machines with sufficient bandwidth and computing power. In our case study (Section~\ref{5-subsec:case-stury-of-UDP}), the attacker can scan the full \texttt{TxID} space in 4.27 seconds on average when the DNS request timeout is 10 seconds.


\noindent \textbf{Practical Considerations.}\label{subsec:DNS_practical_considerations} The feasibility of the attack hinges on three challenges: (1) The spoofed DNS response must precede the legitimate response from the DNS resolver. To enlarge the time window, the attacker can launch a DoS attack against the target DNS resolver by flooding it with a large number of spoofed DNS requests or rerouting its traffic to a black hole by leveraging the ICMP redirect mechanism \cite{DNS-Man-2021, sec21-Tolley, feng2022off-sec-redirect}.
(2) Users often initiate multiple DNS requests simultaneously, creating multiple session entries in the VPN server. The attacker can infer all active sessions in use and inject spoofed DNS responses into each session with fake answers to the target domain name in parallel by manipulating multiple machines (discussed in Section \ref{6-subsec:practical_consideration}). (3) The injection must be completed before the expiration of the DNS request timeout. After our empirical case studies in Section \ref{5-subsec:case-stury-of-UDP}, we confirm that the attacker is frequently able to finish the injection before the timeout.

\section{Real-world Empirical Studies}\label{sec:empirical_study}
\begin{table*}[htb]
\small
\renewcommand\arraystretch{1.1}
\setlength\tabcolsep{1pt}
\centering
\caption{Tested Results of Different Connection Tracking Frameworks.}
\label{Framework_results}
\begin{threeparttable}

\scalebox{0.75}{
\setlength{\tabcolsep}{2mm}{
\begin{tabular}{c|c|c|cc|ccc|c} 
\bottomrule


\multicolumn{3}{c|}{\textbf{System Setup}} 
& \multicolumn{2}{c|}{\textbf{\tabincell{c}{Port Exhausting \\DoS Attack}}} 
& \multicolumn{3}{c|}{\textbf{\tabincell{c}{TCP Hijacking \\Attack}}} 
& \multicolumn{1}{c}{\textbf{\tabincell{c}{DNS Hijacking \\ Attack}}} \\
\hline

\textbf{OS} &  \textbf{\tabincell{c}{Connection \\Tracking \\Framework}}  &  \textbf{\tabincell{c}{Source Port \\Allocation}} 
& \textbf{\tabincell{c}{External Ports\\ Fully Occupiable}} & \textbf{Vulnerable} 
  & \textbf{\tabincell{c}{RST \\ Check}} 
  & \textbf{\tabincell{c}{Expiration \\ Timeout}}   & \textbf{Vulnerable}
  & \textbf{Vulnerable}\\
\hline

\multirow{2}*{Linux} & \multirow{2}*{Netfilter} &preservation
& yes & \blackcheck  & in-window check & 10s & \blackcheck & \blackcheck \\
\cline{3-9}
&  &random & no &  \blackcross  & in-window check & 10s &  \blackcross & \blackcross \\

\hline

\multirow{8}*{FreeBSD} & \multirow{2}*{PF} &preservation & yes  &  \blackcheck & no check & 90s & \blackcheck & \blackcheck \\
\cline{3-9}
 &  &random & yes &  \blackcheck & no check & 90s &  \blackcross& \blackcross \\
\cline{2-9}

 & \multirow{2}*{IPFilter} &preservation& no&  \blackcross & no check & 60s & \blackcheck & \blackcheck \\
\cline{3-9}
 &  &random& no &  \blackcross  & no check &60s &  \blackcross & \blackcross \\
\cline{2-9}

 & \multirow{2}*{IPFW} &preservation& no &  \blackcross& strict check & - & \blackcross & \blackcheck \\
\cline{3-9}
 &  &random& no &  \blackcross& strict check & - &  \blackcross& \blackcross \\
\cline{2-9}

 & \multirow{2}*{natd} &preservation  & yes &  \blackcheck & strict check & - & \blackcross & \blackcheck \\
\cline{3-9}
 & &random & yes &  \blackcheck  & strict check & - &  \blackcross& \blackcross \\
\cline{2-9}

\hline

\toprule
\end{tabular}}
}
\end{threeparttable}
\end{table*}

In this section, we first explore the vulnerabilities in different connection tracking frameworks from Linux and FreeBSD. Then we take empirical case studies to evaluate the effectiveness of the attacks in real-world VPN networks.

\noindent\textbf{Ethical Considerations.}
Our experiments are conducted with careful consideration of ethical problems.
The analyses of different connection tracking frameworks are performed in a fully controlled local environment. For real-world VPN testing, we subscribe to services using two accounts per provider to carry out our non-malicious security research under its safe harbor provisions.
The attacker and victim machines are both under our control. All DNS queries are directed to our own controlled resolvers, and target servers are self-hosted, except in the HTTP injection test, where a public web server is used. However, we only inject fake HTTP responses to our victim client device. 
To minimize any impact on VPN server availability, all traffic during DoS, port inference, and sequence inference phases is rate-limited to no more than 10 MB/s for a duration of 10 seconds. No disruption to VPN services is observed or reported.
We also follow responsible disclosure procedures to notify affected vendors.

\subsection{Connection Tracking Framework Analysis}\label{5-subsec:analysis-of-NAT-frameworks}

We perform tests on open-source connection tracking frameworks from different OSes, i.e., Linux \texttt{Netfilter}, FreeBSD \texttt{PF}, \texttt{IPFW}, \texttt{IPFilter}, and \texttt{natd}, as they may be chosen by VPN providers.
For each framework, we test if it violates the non-interference property and may be vulnerable to the attacks under different configurations. The results are shown in Table \ref{Framework_results}.


\noindent\textbf{Port Exhausting DoS Attack.} We find that attackers can exhaust all usable external ports in \texttt{Netfilter} (with \textit{Port Preservation}), \texttt{PF}, and \texttt{natd} under both port allocation strategies, while the attack is ineffective against \texttt{IPFilter} and \texttt{IPFW} due to inherent session entry limits.
For \texttt{Netfilter} with \textit{Port Preservation}, an attacker can occupy all external ports to the target server, causing victim packets to be dropped. However, under \textit{Random Selection}, it becomes increasingly hard for the port selection algorithm to find unused source ports to create new session entries, making it difficult for an attacker to exhaust all ports.
In \texttt{PF}, both allocation strategies are vulnerable. Under \textit{Port Preservation}, an attacker can exhaust all 65,535 external ports. With \textit{Random Selection}, \texttt{PF} selects ports from 50001 to 65535. When these ports become fully occupied, a bug in the implementation causes subsequent packets to be sent with the internal IP address without NAT. This issue has been acknowledged and fixed by the maintainers.
In \texttt{natd}, both strategies are also vulnerable. Under \textit{Random Selection}, the full range from 32768 to 65535 can be consumed. A similar bug to that in \texttt{PF} is triggered afterward, causing packets to be sent with private IPs instead of being NATed.
In contrast, \texttt{IPFilter} limits session entries to 30,000 under \textit{Port Preservation} and 256 under \textit{Random Selection}, while \texttt{IPFW} enforces a strict limit of 16,384 entries under both modes. These constraints effectively prevent port exhaustion attacks in both frameworks.

\noindent\textbf{TCP Hijacking Attack}. The attack can succeed if the framework adopts \textit{Port Preservation} and lacks sufficient verification in state transition when receiving \texttt{RST} packets.
We find \texttt{Netfilter}, \texttt{PF}, and \texttt{IPFilter} with \textit{Port Preservation} are vulnerable.
For \texttt{Netfilter}, if a TCP \texttt{RST} packet with a controlled \texttt{TTL} value has a sequence number within the challenge ACK window (around $2^{16}$), the session entry remains in the \textit{ESTABLISHED} state, but the timeout drops to 10 seconds.
\texttt{PF} and \texttt{IPFilter} are even more vulnerable. They accept \texttt{RST} packets with arbitrary sequence numbers and transition the session to \textit{CLOSE}, lasting 90 seconds in \texttt{PF} and 60 seconds in \texttt{IPFilter}, both shorter than the original \textit{ESTABLISHED} timeout. FreeBSD has assigned a high-severity CVE (CVE-2023-6534) for the problem.
In contrast, \texttt{IPFW} and \texttt{natd} strictly validate \texttt{RST} packet sequence numbers and are not vulnerable to this attack.

%

\noindent\textbf{DNS Hijacking Attack}.
As stated in Section~\ref{subsec:DNS_infer_port}, if the attacker can infer the source port used by the victim's DNS query, it can inject malicious responses into it. Our method becomes feasible when the connection tracking framework employs \textit{Port Preservation}.
As a result, \texttt{Netfilter}, \texttt{PF}, \texttt{IPFilter}, \texttt{IPFW}, and \texttt{natd} with \textit{Port Preservation} are all vulnerable to the attack.

\subsection{Case Study of Port Exhausting DoS Attack}\label{5-subsec:Methodology}

\subsubsection{Experiment Setup}\label{5-subsubsec:DoS-Experiment-Setup}

Our testbed includes four machines: attacker, victim client, and target server (all running Ubuntu 22.04), plus a VPN server selected from commercial providers. In this work, we investigate the feasibility of the attacks on 9 of the top 10 commercial VPN providers \cite{top10vpn}, excluding \textit{Hide.me} due to service unavailability.
For each provider, we choose a VPN server according to the application's suggestion and use OpenVPN as the default protocol.
All machines are geographically separated in different regions to simulate real-world cases and are connected as depicted in Figure \ref{fig:threat-model}.
The victim client is securely configured to route all traffic through the VPN server, following best practices.

\begin{table*}[t]
\small
\renewcommand\arraystretch{1.5}
\setlength\tabcolsep{1pt}
\centering
\caption{Attack Evaluation of Different VPN Providers.}
\label{table:attack_results}
\begin{threeparttable}

\scalebox{0.75}{
\setlength{\tabcolsep}{2mm}{
\begin{tabular}{c|ccc|cc|cc|ccc} 
\bottomrule


\multicolumn{1}{c|}{\textbf{\tabincell{c}{VPN \\Information}}} 
& \multicolumn{3}{c|}{\textbf{\tabincell{c}{Measurement \\Result}}} 
& \multicolumn{2}{c|}{\textbf{\tabincell{c}{Port Exhausting \\DoS Attack}}} 
& \multicolumn{2}{c|}{\textbf{\tabincell{c}{TCP Hijacking \\Attack}}} 
& \multicolumn{3}{c}{\textbf{\tabincell{c}{DNS Hijacking \\ Attack}}} \\
\hline

\textbf{\tabincell{c}{VPN \\Provider}} 

& \textbf{\tabincell{c}{Session\\ Limit}} 
& \textbf{\tabincell{c}{Source Port \\ Assignment}} 
& \textbf{\tabincell{c}{RST \\ Check}}

& \textbf{\tabincell{c}{TCP \\ DoS}} 
& \textbf{\tabincell{c}{DNS \\ DoS}} 
& \textbf{\tabincell{c}{HTTP \\ Injection}}   
& \textbf{\tabincell{c}{FTP \\Hijacking}} 
& \textbf{\tabincell{c}{5s \\timeout}}  
& \textbf{\tabincell{c}{10s \\timeout}}
& \textbf{\tabincell{c}{15s \\timeout}} \\
\hline


ExpressVPN  & no & preservation & in-window, 10s & $4.14 | \frac{10}{10}$ & N/V & $74.62 | \frac{6}{10}$ & $21.45 | \frac{8}{10}$ & N/V & N/V & N/V\\

\hline


NordVPN & \tabincell{c}{no \\ yes} & \tabincell{c}{preservation \\ random} & \tabincell{c}{in-window, 10s \\ strict-check}  & N/V & N/V & N/V & $86.75 | \frac{7}{10}$ & N/V & N/V & N/V\\

\hline
PIA & no & preservation & in-window, 10s & $3.97 | \frac{10}{10}$ & $4.30 | \frac{10}{10}$ & $64.11 | \frac{7}{10}$ & $22.77 | \frac{8}{10}$ & $4.74 | \frac{2}{10}$ & $7.17 | \frac{5}{10}$ & $8.53 | \frac{7}{10}$ \\

\hline
Surfshark & \tabincell{c}{no \\ yes} & \tabincell{c}{preservation \\ random} & \tabincell{c}{in-window, 10s \\ strict-check}  & N/V & N/V & N/V & $24.70 | \frac{8}{10}$ & N/V & N/V & N/V\\


\hline
PrivateVPN & no & preservation & in-window, 10s & $4.36 | \frac{10}{10}$ & $4.05 | \frac{10}{10}$ & $58.45 | \frac{7}{10}$ & $21.38 | \frac{8}{10}$ & $4.40 | \frac{2}{10}$ & $7.16 | \frac{6}{10}$ & $8.48 | \frac{7}{10}$\\

\hline
IPVanish & no & preservation & in-window, 10s & $3.98 | \frac{9}{10}$ & N/V & $67.71 | \frac{7}{10}$ & $23.44 |\frac{9}{10}$ & N/V & N/V & N/V\\

\hline
CyberGhost & no & preservation & in-window, 10s & $3.99 | \frac{10}{10}$ & $4.31 | \frac{9}{10}$ & $67.48 | \frac{7}{10}$ & $22.57 | \frac{9}{10}$ & $4.86 | \frac{3}{10}$ & $7.11 | \frac{6}{10}$ & $7.72 | \frac{6}{10}$ \\


\hline

ProtonVPN & yes & random & strict-check & N/V & N/V & N/V & N/V & N/V & N/V & N/V\\

\hline

Windscribe   & no & preservation & in-window, 10s & $3.99 | \frac{10}{10}$ & $4.28 | \frac{9}{10}$ & $63.69 | \frac{6}{10}$ & $26.63 | \frac{8}{10}$ & $4.95 | \frac{1}{10}$ & $8.45 | \frac{5}{10}$ & $8.54 | \frac{8}{10}$\\

\hline

\toprule
\end{tabular}}
}
   \begin{tablenotes}
       \footnotesize
       \item[*] N/V means that the VPN server is \textbf{Not Vulnerable} to this attack.
    \end{tablenotes}
\end{threeparttable}
\end{table*}

\noindent\textbf{Attack Procedure}. We have two experiments in this attack. First, the target server runs an HTTPS webpage on TCP port 443 using Nginx. The attacker will then occupy all the source ports of the VPN server to the target server, following Figure \ref{fig:port_dos}. The victim then attempts to access the web server. 
Second, we configure the target server as the victim's DNS resolver and have the attacker occupy all source ports to it. If the victim cannot resolve domains anymore, the attack is deemed successful.
For each VPN provider, we repeat the attacks 10 times and record the time cost and success rate.

\subsubsection{Experiment Results}

Table~\ref{table:attack_results} summarizes the attack results, listing time cost (in seconds) and success rate. 
For instance, the column of TCP DoS attack on ExpressVPN (i.e., $4.14 | \frac{10}{10}$) shows success in 4.14 seconds with a 100\% rate.

For TCP DoS, we can successfully exhaust all ephemeral ports of VPN servers from ExpressVPN, PIA, PrivateVPN, IPVanish, CyberGhost, and Windscribe within nearly 4 seconds and effectively obstructed the TCP connections directed to the target server with success rates over 90\%.  
NordVPN, Surfshark, and ProtonVPN deploy proxy-like technologies to manage outgoing traffic~\cite{nord-surfshark-proxy,nord-patent,protonVPN-proxy,protonVPN-accelerator}. Specifically, NordVPN and Surfshark apply the proxy mechanism only to popular destination ports such as 443 and 80, while other ports (e.g., FTP port 21) are not covered. In contrast, ProtonVPN applies proxying to all destination ports. We find that this proxy-based design can effectively prevent the port exhaustion attack.

Similarly, the DNS DoS attack succeeded against PIA, PrivateVPN, CyberGhost, and Windscribe, disrupting resolution in ~4.24 seconds with over 90\% success.
ExpressVPN, NordVPN, Surfshark, and IPVanish have special DNS settings that the victim's DNS queries will not be forwarded to the target server after setting the DNS resolver to it, but the client can still receive the DNS responses with a source IP address of the target server. While this setup protects against the attack, it can cause usability issues if users want to resolve customized domains with its own DNS resolver. ProtonVPN extends its proxy-like protections to UDP, fully blocking the attack.

\subsection{Case Study of TCP Hijacking Attack}\label{5-subsec:case-stury-of-TCP}

\subsubsection{Experiment Setup}


The experimental environment is consistent with that in Section~\ref{5-subsubsec:DoS-Experiment-Setup}. We conduct two case studies: HTTP injection and FTP hijacking.
For HTTP injection, we avoid scanning the entire Internet due to scale and ethical concerns. As reported by~\cite{https-adoption}, approximately 13.9\% of websites still do not default to HTTPS, leaving a non-trivial attack surface. Following prior works~\cite{cao2016off, ccsfeng}, we select a well-known financial website, \textit{{ANONYMOUS}.com}, as our target.
For FTP hijacking, we set our target server to provide file-downloading services through FTP (with \texttt{vsftpd} 3.0.3). 

\noindent\textbf{Attack Procedure}. 
For HTTP injection, the victim's browser maintains a long-lived HTTP session that periodically issues requests to fetch exchange rates. The attacker tries to inject forged responses with fake financial data with our method. For FTP hijacking, the victim logs into the FTP server and issues different FTP commands. The attacker will try to steal the victim's private files during the attack. For each VPN provider, we perform both attacks 10 times independently, renewing the session between the client and server after each run.


\subsubsection{Experiment Results}
\noindent \textbf{VPN Analysis.}
8 VPN providers are vulnerable under our threat model. ProtonVPN is unaffected due to its use of proxy technology, which employs \textit{Random Selection} and strict TCP \texttt{RST} validation. NordVPN and Surfshark mitigate HTTP injection via proxying TCP port 80 traffic, but remain vulnerable to FTP hijacking.

\begin{figure}[ht]
	\begin{center}
		\includegraphics[width=1\linewidth]{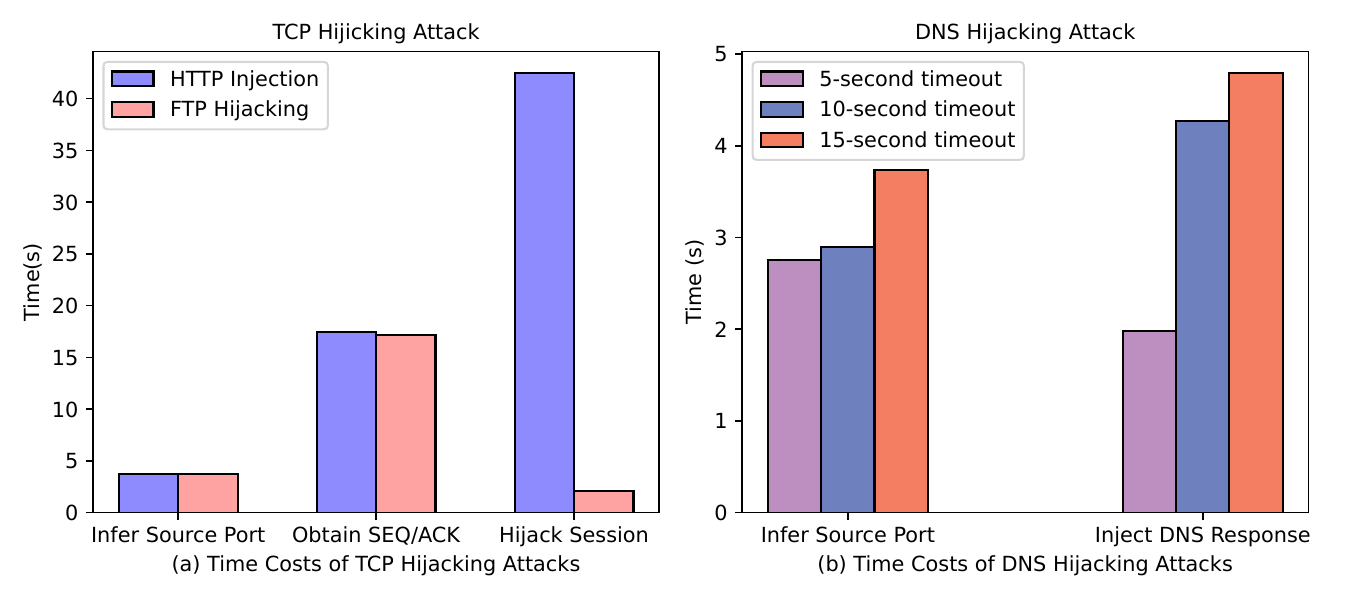}
		\caption{Time cost of the attacks}
		\label{time-cost}
	\end{center}
\end{figure}

\noindent \textbf{Time Cost.}
As time costs across most VPNs follow similar patterns (except NordVPN), we report detailed results in Figure~\ref{time-cost}(a) using PIA as an example.
For HTTP injection, identifying the client's source port takes 3.73 seconds (6.55~MB/s bandwidth), while obtaining sequence numbers takes 17.42 seconds on average. The most time-consuming phase is response injection, averaging 42.43 seconds due to the need to evict the attacker's own session and await the next victim request. The full attack completes in 64.11 seconds on average.
For FTP hijacking, the three phases take 3.69, 17.15, and 2.11 seconds, respectively, resulting in a time cost of 22.77 seconds for the entire attack to get a private file from the server. This attack is faster as it does not require waiting for a new client request.
Notably, NordVPN experiences high packet loss during the inference period. This leads to multiple retries and extended delays in determining the source port, resulting in increased time costs and a lower success rate.
%
%

\begin{figure}[ht]
	\vspace{-1mm}
	\begin{center}
		\includegraphics[width=0.7\linewidth]{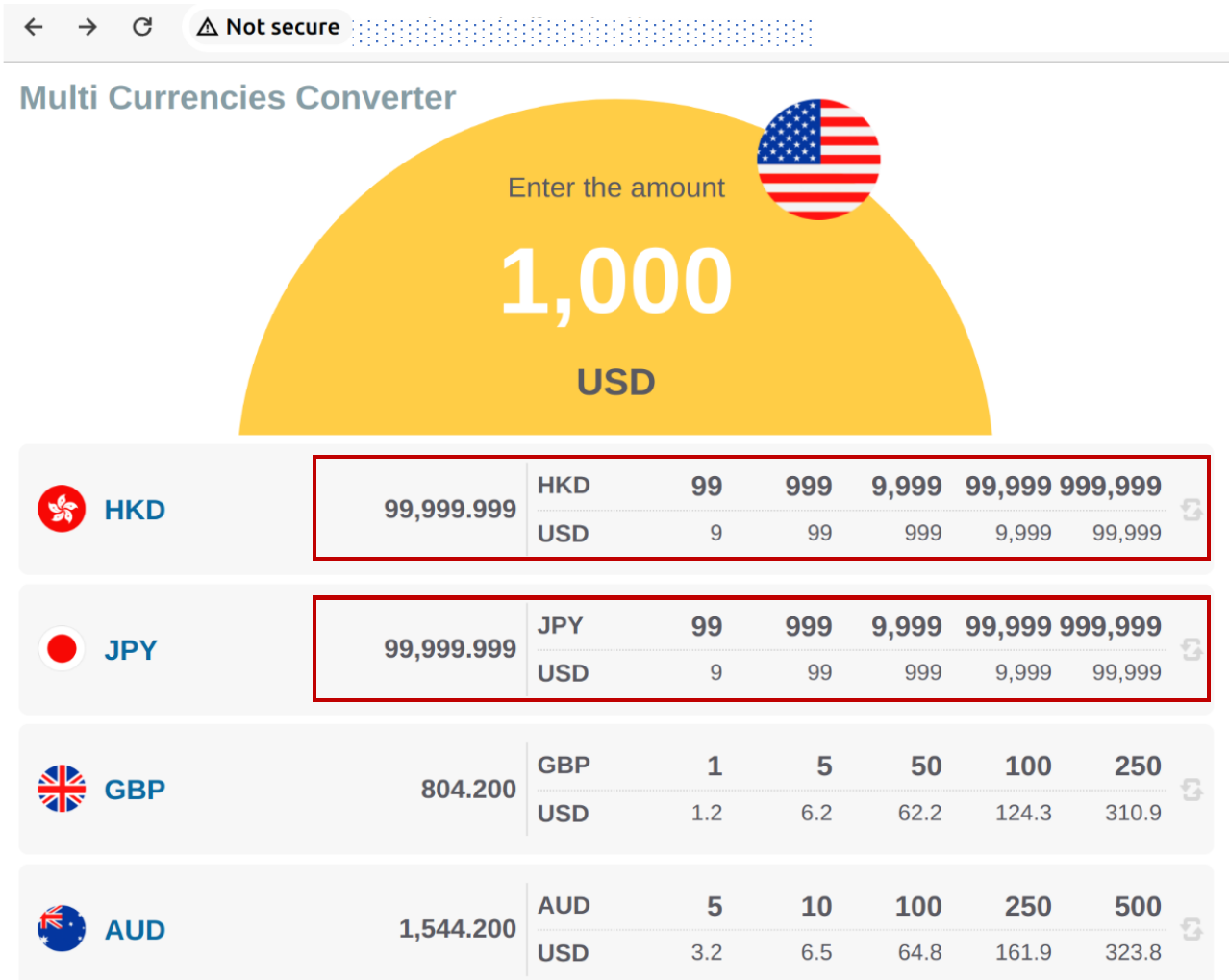}
		\caption{The exchange rates on the web page after attack}
		\label{aastocks}
	\end{center}
\end{figure}

\noindent \textbf{Success Rate}.
For VPN providers where attack conditions are met (ExpressVPN, PIA, PrivateVPN, IPVanish, CyberGhost, Windscribe), HTTP injection succeeds with an average rate of 66.7\%. Failures arise primarily from: (1) packet loss during probing; (2) the client receives valid data from the real server before injection, or (3) as the attacker will temporarily replace the original session entry, if the client sends a request during this period, the server will respond with an \texttt{RST} to terminate the connection.
Figure \ref{aastocks} shows a snapshot of the attack result, in which the exchange rates are manipulated to other forged values and may lead to wrong purchase or sale. 
FTP hijacking achieves an 81.3\% success rate across 8 providers. Failures are mainly due to (1) packet loss and (2) high-frequency client communication, which refreshes session timeouts and prevents entry eviction. We further analyze this in Section~\ref{6-subsec:practical_consideration}.


\subsection{Case Study of DNS Hijacking Attack}\label{5-subsec:case-stury-of-UDP}

\subsubsection{Experiment Setup}

The setup of the experiment is also the same as that in Section \ref{5-subsubsec:DoS-Experiment-Setup}.
To launch the DNS hijacking attack, the attacker needs to first mute the DNS resolvers to preclude premature DNS responses. To avoid impacting other users and raising ethical concerns, we chose to set the victim's DNS resolver to our own server.

\noindent\textbf{Attack Procedure}. We follow a similar methodology as Tolley's work~\cite{sec21-Tolley}. Since forged DNS responses must arrive before query timeouts, which vary by OSes and browsers (5, 10, or 15 seconds), we ran 10 tests for each timeout to evaluate the attack.
In each test, the victim VPN client issues a single DNS lookup using \texttt{nslookup} with a specified timeout. Half a second later, the attacker starts the injection script to infer the DNS source port and inject responses with varying \texttt{TxIDs}.


\subsubsection{Experiment Results}

\noindent \textbf{VPN Analysis.}
As stated in the DNS DoS experiments, ExpressVPN, NordVPN, Surfshark, and IPVanish employ special DNS settings, which prevent the attack. ProtonVPN adopts \textit{Random Selection}, also rendering it safe. In contrast, PIA, PrivateVPN, CyberGhost, and Windscribe are vulnerable to inference of DNS sessions.

\noindent \textbf{Time Cost and Success Rate}.
Table~\ref{table:attack_results} summarizes the attack performance across VPNs. We illustrate the time cost of PIA in Figure \ref{time-cost}(b) as the others share similar trends with it.
For a 5-second DNS timeout (e.g., Edge on Windows 11), it took on average 2.76 seconds to infer the source port and 1.98 seconds to inject the correct \texttt{TxID}, using 2.43 MB/s bandwidth, totaling 4.74 seconds with a 20\% success rate. For the 10-second timeout (e.g., Firefox on Ubuntu 20.04), the attacker can succeed with an average time cost of 2.90 seconds and 4.27 seconds, respectively, and a success rate of 50\%. 
For the 15-second timeout (e.g., Safari on macOS 14.1.2), the attacker can succeed with an average success rate of 70\%, and the time costs of the two steps are 3.74 and 4.79 seconds. 
The attacker can achieve a higher success rate with a longer DNS query timeout, as it has more time to infer the correct source port and scan through the possible \texttt{TxID} space before the UDP socket is closed by the victim client.

\section{Discussion and Countermeasures}\label{sec:countermeasures}
\subsection{Real-world Considerations}\label{6-subsec:practical_consideration}

\noindent \textbf{Impacts of Multiple Sessions.} 
Modern browsers often establish multiple concurrent TCP sessions to speed up page loading. However, these additional sessions are usually short-lived and exert minimal impact on deducing the targeted long-lived TCP connection. 
It may be possible that the client maintains multiple long-lived TCP sessions with the server, and the same obstacle exists in the DNS hijacking attack. For example, the attacker attempts to inject forged DNS responses for \texttt{a.com}, but meanwhile the client also queries for \texttt{b.com}. 
The attacker can infer the source ports of all active sessions and inject responses into the inferred sessions. We have tested the DNS hijacking attack when the client initiates 10 different DNS queries with a 10-second timeout. The attacker controls 5 machines to perform inference and injection in parallel, and the attack succeeds in 4 of the 10 tests. 

\noindent \textbf{Impacts of Different Traffic Frequencies.}
In TCP hijacking attacks, the attacker must first remove the victim's session entry at the VPN server, which typically expires after a 10-second timeout. However, if the server receives packets matching the session, the timeout is refreshed, forcing the attacker to retry. 
To assess this effect, we evaluate the FTP hijacking attack under varying client-server communication intervals (e.g., 4, 8, 12 seconds). Each configuration is tested 10 times.
Results show that when the interval is below 10 seconds, the attack consistently fails due to the session entry being refreshed. For intervals exceeding 10 seconds, the attack succeeds with an average success rate of 85\%. As the interval increases, the time cost decreases, since the attacker requires fewer attempts and less waiting time.

\subsection{Countermeasures}

\noindent\textbf{Responsible Disclosure.}
We have disclosed the vulnerabilities to Linux, FreeBSD, and affected VPN providers. FreeBSD has released an announcement confirming the vulnerabilities and assigned a high-severity CVE. Maintainers of Linux \texttt{Netfilter} also acknowledge the possible attack and are discussing with us about the patches. They suggest configuring proper firewall rules to mitigate the attack. Six VPN vendors have confirmed the vulnerability and rewarded us with their bug bounty policies. We are working with them to mitigate the attacks. The other 2 VPN vendors are still investigating the vulnerabilities.
In total, our responsible disclosure has resulted in 19 assigned CVE/CNVD identifiers (specifically CVE-2023-6534, CVE-2024-50751 to CVE-2024-50764, and CNVD-2025-\{05945, 06966, 06970, 07025\}). We also recommend our countermeasures to prevent
the attacks.

\noindent\textbf{Limiting Concurrent Sessions.}
The port exhaustion DoS stems from attackers monopolizing the VPN server's shared source ports. We recommend VPN providers limit concurrent sessions to the same target by leveraging connection tracking limits~\cite{Netfilter-connlimit}, setting firewall rules to curb port scanning, or deploying DDoS protections like SYN-proxy, thereby preventing exhaustion of session entries.


\noindent\textbf{Randomizing Port Allocation.}
It is recommended for the VPN servers to adopt \textit{Random Selection}, which will preclude the attacker from deducing used ports through the port preservation side channel employed in our attacks, thereby obviating the TCP and DNS hijacking attacks. Specifically, upon creating new entries to record sessions, the VPN server can select a random unused source port and record the port translation in the session entries. 

\noindent\textbf{Enforcing Strict RST Checks.} 
As the vulnerable connection tracking frameworks falsely transfer the states of session entries upon receiving illegal \texttt{RST} packets, it is also required to strictly check \texttt{RST} packets such that only those with exact sequence numbers can cause state transition of the session entries, thereby preventing attackers from removing the session entries intentionally to mitigate TCP hijacking attacks. FreeBSD has quickly released patches to fix the vulnerability with this suggestion. However, the maintainers of Linux Netfilter argue that it's hard for middleboxes to always track the states or values of endpoints. They are concerned that strict checks of \texttt{RST} packets may prevent the connection tracking table entries from being properly cleared, leading to resource exhaustion. We are still working with them to promote better patches to be applied in the kernel.


\FinalRevise{
\section{Related Work}\label{sec:relatedwork}
\noindent \textbf{VPN Security.}
The security of VPN systems has been extensively investigated across platforms such as Android, iOS, and desktop systems~\cite{VPN-alyzer-ndss22, vpn-configuration-imc16, vpn-ios-investigate-2020}, revealing issues such as traffic leakage and misconfiguration.
Perta et al.~\cite{vpn-dns-hijack-pets15} and Tolley et al.~\cite{sec21-Tolley} demonstrated DNS and session hijacking attacks via malicious access points or in/on-path adversaries. Xue et al.~\cite{vpn-routetable-sec23} and Moratti et al.~\cite{tunnel-vision} exploited routing table manipulation to divert VPN traffic.
More recently, Mixon-Baca et al.~\cite{MixonBaca-pets24} introduced the \textit{Port Shadow} attack, which exploits port collisions between a client's source port and the VPN server's listening port to infer VPN usage. In contrast, our work targets \textit{client-to-external-server} sessions and exploits port collisions between VPN clients sharing the same connection tracking table. Furthermore, we uncover a distinct vulnerability in the handling of TCP \texttt{RST} packets that results in incorrect session state transitions, enabling effective inference and hijacking attacks. Moreover, our work introduces different attack effects (e.g., DoS). Unlike \textit{Port Shadow}, our attack does not assume that the VPN server shares the same entry and exit IP address, making it applicable in more common VPN deployments.

\noindent \textbf{NAT Security.}
Previous studies have investigated NAT behaviors for host enumeration~\cite{NAT-by-ipid,NAT-by-port,NAT-by-clock}. Winemiller et al.~\cite{winemiller2012nat} and Nguyen et al.~\cite{slowdos-CANDAR-2018,slowdos-2022} proposed NAT DoS attacks that fill all the NAT table in hypervisors or Docker containers via a compromised machine. Feng et al.~\cite{fengredan} demonstrated remote TCP DoS attacks against NAT networks. These attacks typically rely on predictable identifier fields or compromised internal hosts.
Gilad et al.~\cite{gilad-fragmentation-tcp-2013} showed that IP fragmentation can be used to bypass NATs for interception, while Herzberg et al.~\cite{dns-Herzberg-tpds12} inferred DNS source ports by pre-filling NAT tables using puppets. However, their method targets outdated NAT implementations.
In contrast, our work targets VPN environments where connection tracking frameworks are shared among tenants. We exploit port collision behavior that arises \textit{after} the victim session is established, without requiring puppets. Moreover, our selective session-level DoS does not rely on flooding the table, making it more stealthy.

\noindent \textbf{Session Manipulation Attacks.}
TCP hijacking attacks have leveraged side channels such as challenge ACKs~\cite{cao2016off}, IPID behaviors~\cite{ccsfeng}, timing patterns~\cite{chen2018off,qian2012}, and packet sizes~\cite{wang2025off}. Yang et al.~\cite{yangexploiting, yangton} exploited the router vulnerability that totally disabled TCP window tracking, enabling TCP hijacking in Wi-Fi environments. Feng et al.~\cite{feng2023man-sp-wifi-redirect} leveraged ICMP redirects to conduct Man-in-the-Middle attacks in Wi-Fi networks. Our work targets stricter VPN environments where connection tracking enforces state validation and reverse path checks. Furthermore, we uncover attacks without proximity to the victim and go beyond TCP hijacking to include port exhaustion DoS and DNS injection.
For DNS manipulation, Man et al.~\cite{DNS-Man-2020,DNS-Man-2021} inferred DNS ports via ICMP side channels, while Herzberg et al. and Zheng et al.~\cite{herzberg-fragmentation-dns-2013-cns,DNS-Zheng-2020} used IP fragmentation for cache poisoning. In contrast, our method reveals a novel port collision channel within VPN servers, allowing a malicious VPN user to infer DNS source ports and mount hijacking attacks without network-level access or fragmentation.
}
\section{Conclusion}\label{sec:conclusion}

In this work, we uncover new off-path session manipulation attacks in VPN networks that malicious users can abuse the vulnerabilities that exist in the shared connection tracking table and improper state transition of session entry, which violates the non-interference property required for the VPN servers. We present our attacks on 5 mainstream connection tracking frameworks and 9 commercial VPN providers and demonstrate that off-path attackers can disrupt or hijack other client's TCP and DNS sessions. We confirm the vulnerabilities in all frameworks and 8 VPN providers. Our findings pose new challenges to the current understanding of real-world VPN's security and have led to 19 assigned CVEs/CNVDs through responsible disclosure. The authors have provided public access to their code at \url{https://github.com/yyxRoy/vpn-attacks}.

\section*{Acknowledgment}
The work is in part supported by the National Science Foundation for Distinguished Young Scholars of China under Grant 62425201, the Science Fund for Creative Research Groups of the National Natural Science Foundation of China under Grant 62221003, and the National Natural Science Foundation of China under Grant 62132011. Ke Xu is the corresponding author.


\bibliographystyle{IEEEtran}
\bibliography{ref}

\end{document}